\documentclass[%
 reprint,
 amsmath,amssymb,
 aps,
 pra,
]{revtex4-2}

\usepackage{graphicx}
\usepackage{bm}
\usepackage{xcolor}
\usepackage[title]{appendix}

\begin{document}


\title{Stochastic Representation of Time-Evolving Neural Network-based Wavefunctions}

\author{Bizi Huang}
 \email{huangbizi@stu.pku.edu.cn}
 \affiliation{School of Physics, Peking University, Beijing 100871, People's Republic of China}

\author{Weizhong Fu}
 \email{fuwz@pku.edu.cn}
 \affiliation{School of Physics, Peking University, Beijing 100871, People's Republic of China}

\author{Ji Chen}
\email{ji.chen@pku.edu.cn}
\affiliation{School of Physics, Peking University, Beijing 100871, People's Republic of China}
\affiliation{Interdisciplinary Institute of Light-Element Quantum Materials and Research Center for Light-Element Advanced Materials, Peking University, Beijing 100871, People's Republic of China}
\affiliation{State Key Laboratory of Artificial Microstructure and Mesoscopic Physics, Frontiers Science Center for Nano-Optoelectronics, Peking University, Beijing 100871, People's Republic of China}


\begin{abstract}
Solving the time-dependent Schrödinger equation (TDSE) is pivotal for modeling non-adiabatic electron dynamics, a key process in ultrafast spectroscopy and laser-matter interactions. 
However, exact solutions to the TDSE remain computationally prohibitive for most realistic systems, as the Hilbert space expands exponentially with dimensionality.
In this work, we propose an approach integrating the stochastic representation framework with a neural network wavefunction ansatz, a flexible model capable of approximating time-evolving quantum wavefunctions. 
We first validate the method on one-dimensional single-electron systems, focusing on ionization dynamics under intense laser fields, a critical process in attosecond physics.
Our results demonstrate that the approach accurately reproduces key features of quantum evolution, including the energy and dipole evolution during ionization. 
We further show the feasibility of extending this approach to three-dimensional systems. Due to the increased complexity of real-time simulations in higher dimensions, these results remain at an early stage and highlight the need for more advanced stabilization strategies.
\end{abstract}

\maketitle

\section{Introduction}

For small and slowly varying external perturbation fields, electrons can be treated as adiabatically evolving between the ground and excited states. Under the adiabatic approximation, methods based on linear response theory have been developed to obtain various properties of excited states~\cite{physchem.55.091602.094449,CASIDA20093,li2005time}. However, when the magnitude of the external perturbation field is comparable to that of the electron binding potential, electrons would deviate from equilibrium. In such cases, the explicit solution of the time-dependent Schrödinger equation (TDSE) becomes necessary to simulate their evolution trajectories. This approach enables the study of non-linear effects, such as high-harmonic generation~\cite{RevModPhys.81.163}, multiphoton ionization~\cite{MORELLEC198297}, and non-equilibrium electron dynamics ~\cite{CORREA2018291,doi:10.1021/jacs.9b10533}. Compared to the time-independent Schrödinger equation, real-time quantum dynamics poses greater theoretical and computational challenges due to the need to explore large regions of the Hilbert space~\cite{guther2018time}.

A key challenge in real-time quantum dynamics is the efficient representation of the time-evolving wavefunction.
The grid method~\cite{LEFORESTIER199159} discretizes the wavefunction on a spatial grid and propagates its evolution via numerical time integration.
Although its accuracy can be systematically improved by refining the grid, the number of grid points grows exponentially with the system's dimensionality.
Moreover, processes like ionization require large simulation domains, and special techniques have been developed to suppress unphysical boundary reflections~\cite{PhysRevA.45.4998, PhysRevA.81.053845, PhysRevA.78.032502}, yet scaling them to high-dimensional systems remains a persistent challenge.
To circumvent the exponential scaling of grid-based methods, an alternative strategy involves representing the wavefunction using parameterized analytical functions~\cite{PhysRevA.55.3417, PhysRevA.88.023422, PhysRevLett.103.063002}.
In recent years, neural networks have emerged as powerful tools for this purpose, achieving notable success in both ground state calculations~\cite{Carleo2017, Gao2017, Glasser2018, Luo2019, Hermann2020, Pfau2020, Spencer2020, ren2023towards, li2022ab, fu2024variance, fu2025local} and real-time wavefunction evolutions~\cite{Carleo2017, carleo2017unitary, gutierrez2022real, sinibaldi2023unbiasing, Nys2024, sinibaldi2024time, van2024many}.
By combining expressive neural network wavefunctions with the variational principle, or the time-dependent variational principle (TDVP) for real-time dynamics, highly accurate descriptions of ground states and dynamical evolution can be achieved.
However, the large number of parameters of neural networks also leads to substantial computational cost.

Recently, another strategy, namely the stochastic representation of the wavefunction, has demonstrated remarkable success in imaginary-time evolution~\cite{Atanasova2023,bernheimer2024determinant}.
This method involves fitting the wavefunction to a set of stochastically sampled walkers as they evolve, rather than directly optimizing wavefunction parameters via the variational principle.
Within the framework of stochastic representation, explicit determinant-based constructions for enforcing exchange (anti)symmetry are no longer necessary.
Alternatively, the model learns this property directly from training data augmentation by permuting particles coordinates, thereby substantially reducing computational costs~\cite{Atanasova2023}.
%
That said, whether the stochastic representation strategy can adapt to real-time dynamics remains an open question, as simulating real-time evolution is inherently much more complex than imaginary-time evolution.

In this work, we extend the stochastic representation framework to real-time propagation for solving the TDSE using neural network-based wavefunctions.
We evaluate its performance on one-dimensional and three-dimensional systems exposed to intense femtosecond laser pulses.
%
Unlike conventional wavefunction optimization methods, our approach combines stochastic sampling with evolution-based refinement, directly learning the evolved wavefunction values on sampled points.
This enables efficient, scalable simulations of quantum dynamics.

\section{Methods}
\subsection{Real-time Propagation and TDVP}
The TDSE in a time-varying laser field is given by:
\begin{equation}
\label{eq:tdse}
i \frac{\partial}{\partial t} \psi(\mathbf{r}, t) = \left[ -\frac{1}{2} \nabla^2 + V(\mathbf{r}) + V_{\mathrm{ext}}(\mathbf{r}, t) \right] \psi(\mathbf{r}, t),
\end{equation}
where $\mathbf{r}$ denotes the electron coordinates and $V_{\mathrm{ext}}(\mathbf{r}, t)$ represents the external laser field.
For function-based methods~\cite{PhysRevA.55.3417, PhysRevA.88.023422, PhysRevLett.103.063002} developed to solve the TDSE, the wavefunction's time trajectory is generally determined in parameter space based on TDVP~\cite{Dirac_1930}.
The evolution of the wavefunction parameters is obtained by minimizing residual loss, expressed as:
\begin{equation}
\dot{q} = \arg\min_{\dot{q}} \left\| \hat{H} \psi_q - i\sum_j \frac{\partial \psi_q}{\partial q_j} \dot{q}_j \right\|_{L^2},
\end{equation}
where $q$ represents the parameters of the wavefunction ansatz.
The optimal parameter evolution corresponds to the orthogonal projection of the exact time derivative onto the tangent space of the variational manifold, leading to the orthogonality condition:
\begin{equation}
\left\langle \frac{\partial \psi}{\partial q_j} \Big| i \dot{\psi} - H \psi \right\rangle = 0.
\end{equation}
This yields the following matrix equations of motion:
\begin{equation}
\label{eq:tdvp_eom}
i M \dot{q} = v, \quad M_{ij} = \left\langle \frac{\partial \psi}{\partial q_i} \Big| \frac{\partial \psi}{\partial q_j} \right\rangle, \quad v_i = \left\langle \frac{\partial \psi}{\partial q_i} \Big| H \psi \right\rangle.
\end{equation}
However, numerical instabilities arise not only from finite time-step errors but also from ill-conditioned $M$ matrices, especially when using high-capacity, nonlinear ansatzes such as neural networks.
This issue has been documented in diverse quantum systems, including quantum lattice models~\cite{haegeman2016}, one-dimensional tunneling dynamics~\cite{kvaal2022no}, quantum circuits~\cite{PhysRevResearch.3.033083}, and neural quantum states~\cite{schmitt2025simulating}.
The time-dependent variational Monte Carlo method (tVMC)~\cite{carleo2017unitary,Nys2024} employs Monte Carlo methods to estimate the matrix elements in Eq.~\eqref{eq:tdvp_eom} for neural network wavefunctions.
Nevertheless, as the nonlinearity of the ansatz increases, the equation becomes increasingly stiff.
While the projected variant of tVMC (p-tVMC)~\cite{sinibaldi2023unbiasing} improves numerical stability through implicit optimization and high-order approximations to unitary evolution, it still suffers from two key limitations: low sampling efficiency and accumulation of stochastic variance over time~\cite{Gravina2025neuralprojected}.





\subsection{Stochastic Representation}

\begin{figure}
    \centering       
    \includegraphics[width=0.45\textwidth]{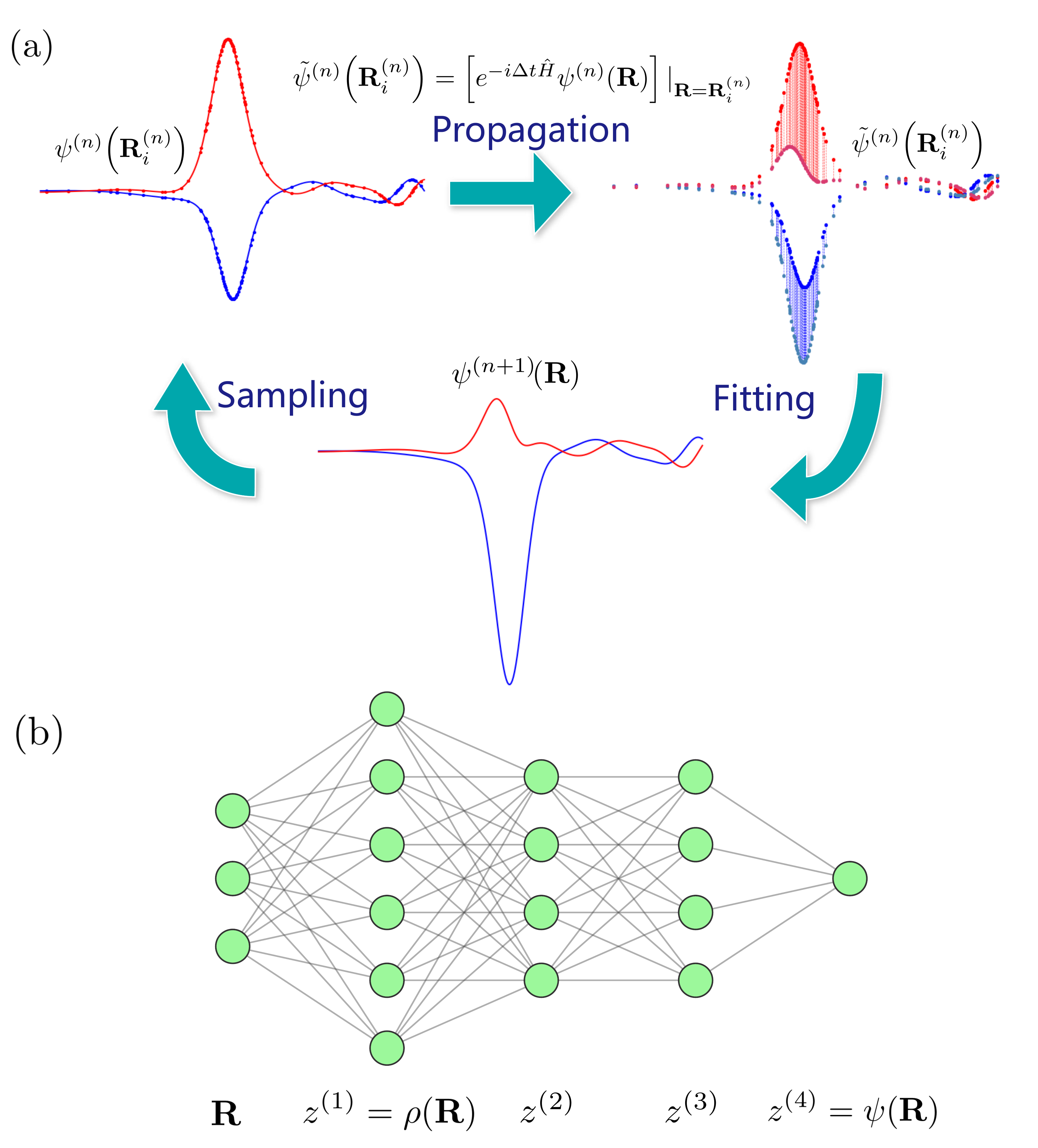}  
    \caption{{\bf Workflow and RBF neural network architecture.} (a) Workflow of real-time propagation with stochastic representation. Each iteration consists of three key steps: first, the RBF neural network wavefunction is represented using stochastic samples; second, the wavefunction values at these sampled positions are propagated under the real-time evolution operator; and third, the RBF neural network is refined by fitting it to the propagated wavefunction values. The red line represents the real part of the wavefunction, and the blue line represents the imaginary part. (b) The RBF neural network architecture. The neural network take the sample coordinates as inputs and outputs the real part or the imaginary part of the wavefunction value $\psi(\mathbf{R})$. The first hidden layer is RBF layer, and the following hidden layers are fully connected layers with tanh activation functions.}
    \label{fig:workflow and RBF network} 
\end{figure}

In the stochastic representation framework, the wavefunction at iteration $n$ is explicitly represented by a set of samples together with their associated wavefunction values $\left\{\left(\mathbf{R}_i^{(n)}, \psi^{(n)}\left(\mathbf{R}_i^{(n)}\right)\right)\right\}$~\cite{Atanasova2023}. This differs fundamentally from diffusion Monte Carlo~\cite{TOULOUSE2016285}, where the wavefunction is represented implicitly through the spatial density of walkers. Under the real-time evolution operator and for a small time step $\Delta t$, each wavefunction sample is propagated as
\begin{equation}
\begin{aligned}
    &\tilde{\psi}^{(n)}\!\left(\mathbf{R}_i^{(n)}\right) 
    = \left[e^{-i\Delta t \hat{H}} \psi^{(n)}(\mathbf{R})\right]|_{\mathbf{R}=\mathbf{R}_i^{(n)}} \\
    &\simeq \psi^{(n)}\!\left(\mathbf{R}_i^{(n)}\right) 
    - i\Delta t \left[\hat{H}\psi^{(n)}(\mathbf{R})\right]|_{\mathbf{R}=\mathbf{R}_i^{(n)}}.
\end{aligned}
\end{equation}
Since the wavefunction generally becomes complex during real-time evolution, it is decomposed into real and imaginary parts as $\psi^{(n)}=u^{(n)}+iv^{(n)}$. 
As illustrated in FIG. \ref{fig:workflow and RBF network}(a), the propagation proceeds through the following steps:
\begin{enumerate}
    \item Obtain the ground state wavefunction (e.g., using the grid method) and fit it with a neural network to obtain $\psi^{(0)}(\mathbf{R})$. Copy parameters to two networks representing the real and imaginary parts, $u^{(0)}(\mathbf{R})$ and $v^{(0)}(\mathbf{R})$.
    \item At iteration $n$, sample the wavefunction at time $n\Delta t$ to generate sample points $\left\{\left(\mathbf{R}_i^{(n)}, \psi^{(n)}\left(\mathbf{R}_i^{(n)}\right)\right)\right\}$.
    \item Alternate time propagation and wavefunction updates:
    \begin{itemize}
        \item Propagate the real part: $\tilde{u}^{(n)}(\mathbf{R}_i^{(n)}) = u^{(n)}(\mathbf{R}_i^{(n)}) + \Delta t \left[\hat{H} v^{(n)}(\mathbf{R})\right]|_{\mathbf{R}=\mathbf{R}_i^{(n)}}$.
        \item Perform supervised learning to train the network for $u^{(n+1)}(\mathbf{R})$ using the propagated samples $\left\{\tilde{u}^{(n)}(\mathbf{R}_i^{(n)})\right\}$.
        \item Propagate the imaginary part: $\tilde{v}^{(n)}(\mathbf{R}_i^{(n)}) = v^{(n)}(\mathbf{R}_i^{(n)}) - \Delta t \left[\hat{H} u^{(n+1)}(\mathbf{R})\right]|_{\mathbf{R}=\mathbf{R}_i^{(n)}}$.
        \item Perform supervised learning to train the network for $v^{(n+1)}(\mathbf{R})$ using the propagated samples $\left\{\tilde{v}^{(n)}(\mathbf{R}_i^{(n)})\right\}$.
    \end{itemize}
    \item Repeat steps 2 and 3 until the simulation is complete.
\end{enumerate}

Here, $u, v$ are represented by radial basis function (RBF) networks and $\tilde{u}, \tilde{v}$ denote the propagated wavefunction values.
The network structures and fitting details are discussed in the next subsection.
Note that we employ a semi-implicit Euler integrator, which offers improved numerical stability and accuracy compared to the forward Euler method, as detailed in Appendix~\ref{sec:inte}.


Regarding the probability distribution $|\Psi(\mathbf{R})|^2$, the sampling procedure begins with Monte Carlo sampling using the Metropolis–Hastings algorithm~\cite{TOULOUSE2016285} to generate \( N \) samples of \( \mathbf{R}_i \), from which we calculate the maximum distance \( R_{\text{max}} \) relative to the origin \( r = 0 \).
To ensure the neural network correctly outputs zero in regions not yet reached by the wavefunction, an additional \( 10\%N \) samples are randomly drawn from the region between \( R_{\text{max}} \) and \( 2R_{\text{max}} \).
In this additional sampling step, if the absolute value of \( u^{(n)}(\mathbf{R}_i) \) or \( v^{(n)}(\mathbf{R}_i) \) is below a threshold \( \varepsilon = 10^{-3} \), the corresponding values are set to zero during training.
%
This truncation step prevents the network from learning spurious non-zero amplitudes in the tail regions, which could otherwise introduce numerical noise and lead to error accumulation over time.
%
To improve training stability and ensure small relative errors in the wavefunction representation, the initial ground-state wavefunction is rescaled such that its value at $r=0$ equals 100.
This scaling operation enhances numerical precision throughout the training process.

\subsection{Neural Network Architecture and Fitting}
\label{sec:rbf}

To simulate ionization effects over a 100 a.u. time range, the electron wavefunction expands from $\pm 5$ a.u. to $\pm 100$ a.u.
Fitting such a wavefunction with a traditional multilayer perceptron (MLP) is highly challenging: without normalizing the input coordinates $x$, gradient vanishing occurs; when normalizing $x$ to the interval $[-1,1]$, the wavefunction variations are compressed into very narrow interval.
%
This results in effectively high-frequency signals in the normalized coordinates, which standard activation functions struggle to represent without resorting to extremely large network parameters, leading to unstable or exploding gradients during training.
To address these issues, we adopt a radial basis function (RBF) network architecture illustrated in Fig.\ref{fig:workflow and RBF network}(b), where a RBF layer is employed as the network's first layer, mapping input coordinates to $M$ localized features centered at trainable positions across the domain.
The output of the first layer is given by
\begin{equation}
z_k^{(1)} = \rho(\mathbf{r}, \mathbf{c}_k), \quad k \in \{1, 2, \dots, M\},
\end{equation}
with the radial basis function defined as
\begin{equation}
\rho(\mathbf{r}, \mathbf{c}_k) = \exp\!\left(- \sum_{i=1}^d \beta_i \,(r_i - c_{k,i})^2\right).
\end{equation}
Here, $\mathbf{r} \in \mathbb{R}^d$ denotes the input coordinate vector with components $r_i$, $\mathbf{c}_k \in \mathbb{R}^d$ is the trainable center of the $k$-th basis function with components $c_{k,i}$, and $\beta_i > 0$ is the learnable decay parameters controlling the width of the basis functions along dimension $i$.
The RBF layer activates neurons only when coordinates are near the centers $c_k$, which ensures localized responses and maintains stable gradient magnitudes during backpropagation.
This localization property mitigates both vanishing and exploding gradient problems, enabling robust training over large spatial domains. 
Subsequent layers are fully connected with $\tanh$ activation functions, 
\begin{equation}
z^{(n)} = W^{(n)}\,\tanh{(z^{(n-1)})} + b^{(n)},
\end{equation}
and the final layer uses no activation, allowing the network to output unbounded real values.
The real and imaginary parts of the wavefunction, $u(\mathbf{R})$ and $v(\mathbf{R})$, are each represented by two separate networks sharing this architecture.
In our experiments, we observed that the approach is sensitive to the exponential parameter 
$\beta$ in the RBF layer.
%
Without constraints, $\beta$ can grow excessively large, producing high-frequency oscillations that destabilize the algorithm. Consequently, we restricted all $\beta$ values to the range $[0.2, 1.0]$.
With this restriction, each RBF neuron is effective for a neighborhood of radius about 2.0. Therefore, the number of RBF centers should be chosen such that every region of radius 2.0 contains at least one center. 
For the 1D numerical experiments in Sec.~\ref{sec:results}, the architecture comprises 4 layers: an initial layer of 256 RBF neurons, followed by two layers of 64 neurons each using tanh activation, and a final output layer for the wavefunction value. For the 3D case, the architecture is identical to the 1D case, with the exception that the first layer is expanded to 4000 RBF neurons.

The loss function is defined as the $L^2$ norm of the prediction error, with an additional regularization term to encourage an approximately uniform spacing of 2.0 a.u. between adjacent RBF centers (based on our experience), thereby promoting consistent resolution across the spatial domain:
\begin{multline}
L^{(n+1)}[u] = \sum_i|u^{(n+1)}(R_i) - \tilde{u}^{(n)}(R_i)|^2 \\+ \lambda \sum^M_{k=1} (|\mathbf{c}_{k+1} - \mathbf{c}_k| - 2)^2.
\end{multline}
Training is performed using the Adam optimizer~\cite{Kingma2014AdamAM}, with the optimizer's state reset at each time step to prevent outdated momentum information from affecting the updated wavefunction.

%
%

An adaptive sampling strategy is employed to minimize the total number of samples while maintaining high fitting accuracy.
The sample set of size $N$ is divided equally into training and validation sets.
The training proceeds for a maximum of 2000 epochs or until the mean absolute error (MAE) of the validation set data falls below the threshold $\delta = 10^{-2}$.
If this threshold is not reached within the maximum number of epochs, an additional $N/10$ samples are added to the dataset, and the network is retrained.
This iterative process continues until the test MAE drops below $\delta$, indicating convergence.
The total number of samples required to achieve convergence in one time step is then used as the initial sample size for the next time step.
This approach enables efficient allocation of computational resources, adapting to the spatial extent and complexity of the wavefunction as it evolves and spreads over time.
We also found that the evolution accuracy is not highly sensitive to the MAE threshold, provided it is not excessively large, as detailed in Appendix~\ref{sec:mae}.

\section{Results and Discussion}
\label{sec:results}
\begin{figure}
    \centering       
    \includegraphics[width=0.45\textwidth]{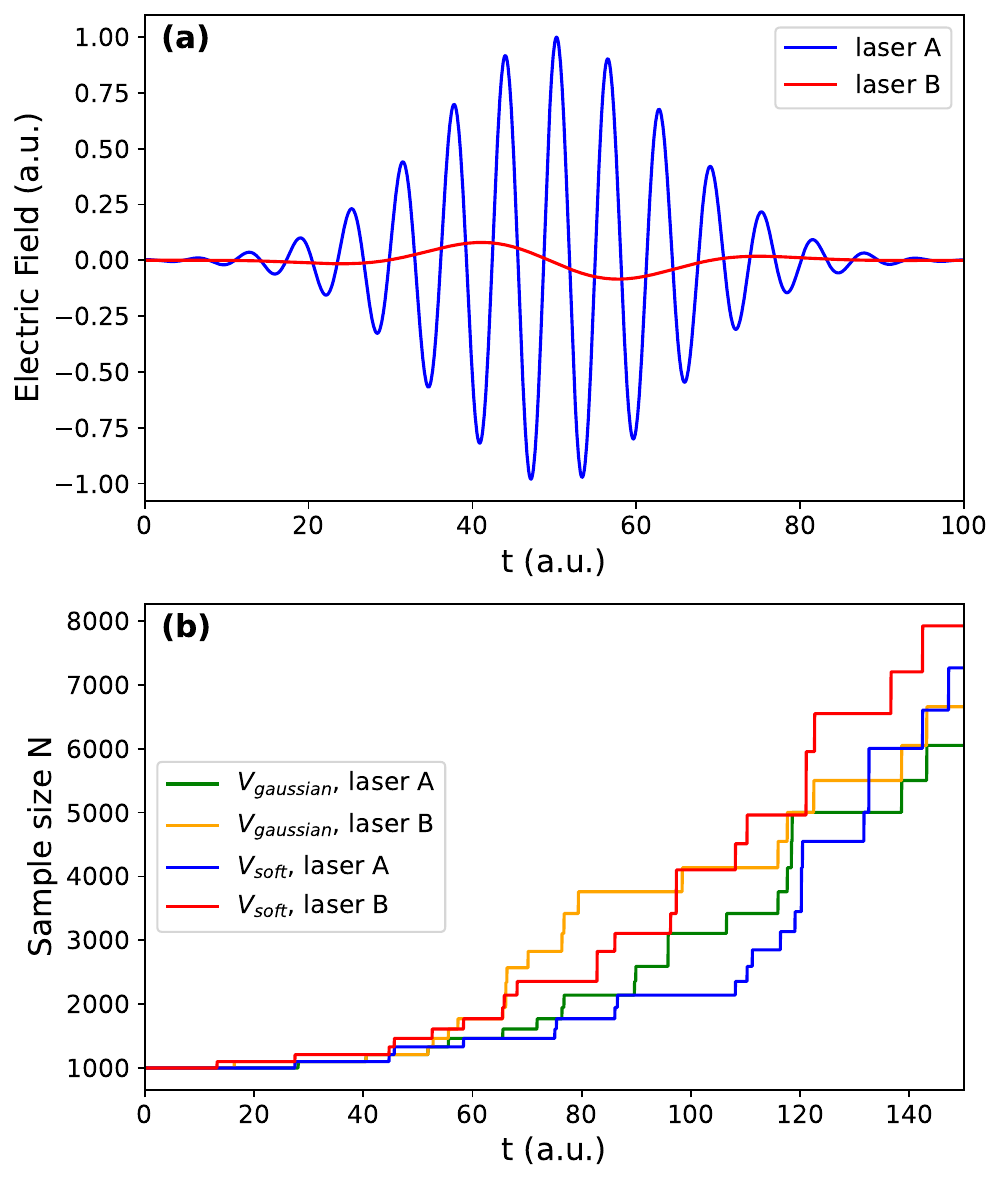}  
    \caption{{\bf Computational details for one-dimensional calculations.} (a) The two different laser fields(units in a.u.): laser A, $E_0=1.0, T =50,\tau=20.5,\omega=1.0$; laser B, $E_0=0.1, T =50,\tau=20.5,\omega=1.0/(2\pi)$. (b) The increasing sample sizes used in the four simulations, showing adaptive growth as required by the convergence criterion.} 
    \label{fig:laser} 
\end{figure}

\begin{figure*}
    \raggedleft
    \noindent
    \includegraphics[width=\linewidth]{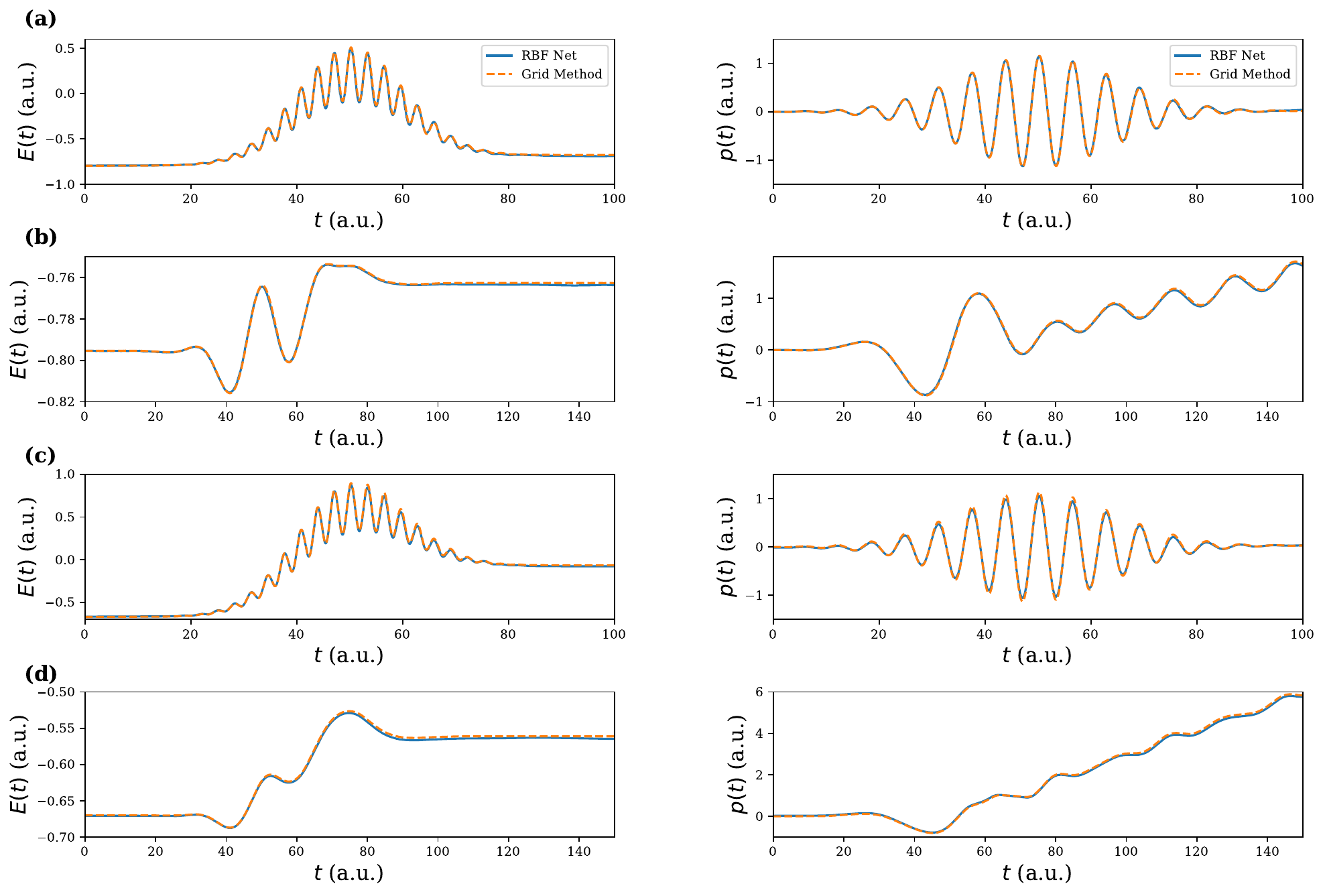}
    \caption{{\bf Simulations of an electron in 1D space under four different conditions.} (a) Gaussian potential with laser A; (b) Gaussian potential with laser B; (c) Soft Coulomb potential with laser A; (d) Soft Coulomb potential with laser B. The left column shows the time evolution of the energy, and the right column shows the time evolution of the dipole moment.}
    \label{fig:1d}
\end{figure*}

We first validate our method using several one-dimensional systems.
Under a time-varying external electric field, the external potential in Eq.~\eqref{eq:tdse} is given by
\begin{equation}
V_{\text{ext}}(\mathbf{r}, t) = -\mathbf{r} \cdot \mathbf{E}(t),
\end{equation}
where $\mathbf{E}(t) = E_0 e^{-(t - T)^2/\tau^2} \cos(\omega t)\mathbf{e}_z$ is the laser electric field, polarized along the $z$-axis.
As displayed in FIG.~\ref{fig:laser}(a), we perform simulations using two different laser pulses with distinct parameters.
Additionally, to avoid the singularity of the potential, we adopt either a Gaussian potential or a soft Coulomb potential instead of the singular Coulomb potential for the electron-nucleus interaction in Eq.~\eqref{eq:tdse}.
These regularized potentials are defined as:
\begin{equation}
    V_{\text{gaussian}}(\mathbf{r}) = -e^{-0.1 r^2}, V_{\text{soft}}(\mathbf{r}) = -\frac{1}{\sqrt{r^2 + 1}}.
\end{equation}
We used the grid method as reliable benchmarks for our simulations, using a grid spacing of $h=0.125$ a.u. and a time step of $\Delta t=0.001$ a.u., which are consistent with the settings in our method.
A total number of 4000 grid points ensures negligible boundary reflection effects in these 1D systems.
%

%

%
FIG.~\ref{fig:laser}(b) illustrates the growth of the sample size in all simulations, reflecting the adaptive sampling strategy employed during training (detailed in Sec.~\ref{sec:rbf}), which dynamically adjusts the number of Monte Carlo samples to maintain accuracy while controlling computational cost. All 1D simulations start with 1000 sample points, far fewer than the 4000 grid points required by the grid method. However, as the wavefunction spreads, the sample size increases to $\sim$8000 to maintain accuracy. This growth is driven by the neural network’s tendency to overfit when trained on sparse training samples, which can generate spurious high-frequency oscillatory components in regions between sampled points. These artifacts increase the validation mean absolute error (MAE), prompting the adaptive sampling mechanism to introduce additional points and restore fidelity. Moreover, these unphysical oscillations lead to local wavefunction regions with artificially large kinetic energy, destabilizing the algorithm. Another source of instability lies in the computation of the Laplacian. Due to overfitting and the lack of gradient information during training, evaluating the Laplacian via automatic differentiation introduces unphysical oscillations. Therefore, rather than following the common practice in neural network–based approaches of using automatic differentiation, we adopt the finite difference method to evaluate the Laplacian, which yields smoother results and improves the stability of our method.

To assess the accuracy of our simulations, the time evolutions of energy and dipole moment are monitored, which are presented in FIG.~\ref{fig:1d}.
For a fair comparison, in the calculations of these observables, we did not use the Monte Carlo sampled points, but employed the same uniform grid points as that of the grid method calculations.
%
%
As shown in FIG.~\ref{fig:1d}, our simulations exhibit excellent agreement with the benchmark results across all test cases.
For laser A, the high-frequency electric field drives rapid oscillations, confining the electron cloud largely to the region around $r=0$.
Laser B has a lower frequency, hence the electric field efficiently excites electrons into scattering states, generating outward-propagating wave packets and a steadily increasing dipole moment. 
Such behavior is particularly difficult to capture with conventional function-based methods, especially those relying on localized basis functions such as standard Gaussians
which struggle to represent delocalized, spreading wavefunctions~\cite{kaufmann1989universal, coccia2016gaussian}.
Despite these challenges, our method performs remarkably well under laser B, accurately capturing the ionization dynamics for both the Gaussian and soft Coulomb potentials.
This demonstrates the robustness and adaptability of the RBF-based stochastic representation in modeling both bound and continuum dynamics over large spatial domains.

\begin{figure}
    \centering
    \includegraphics[width=1.0\linewidth]{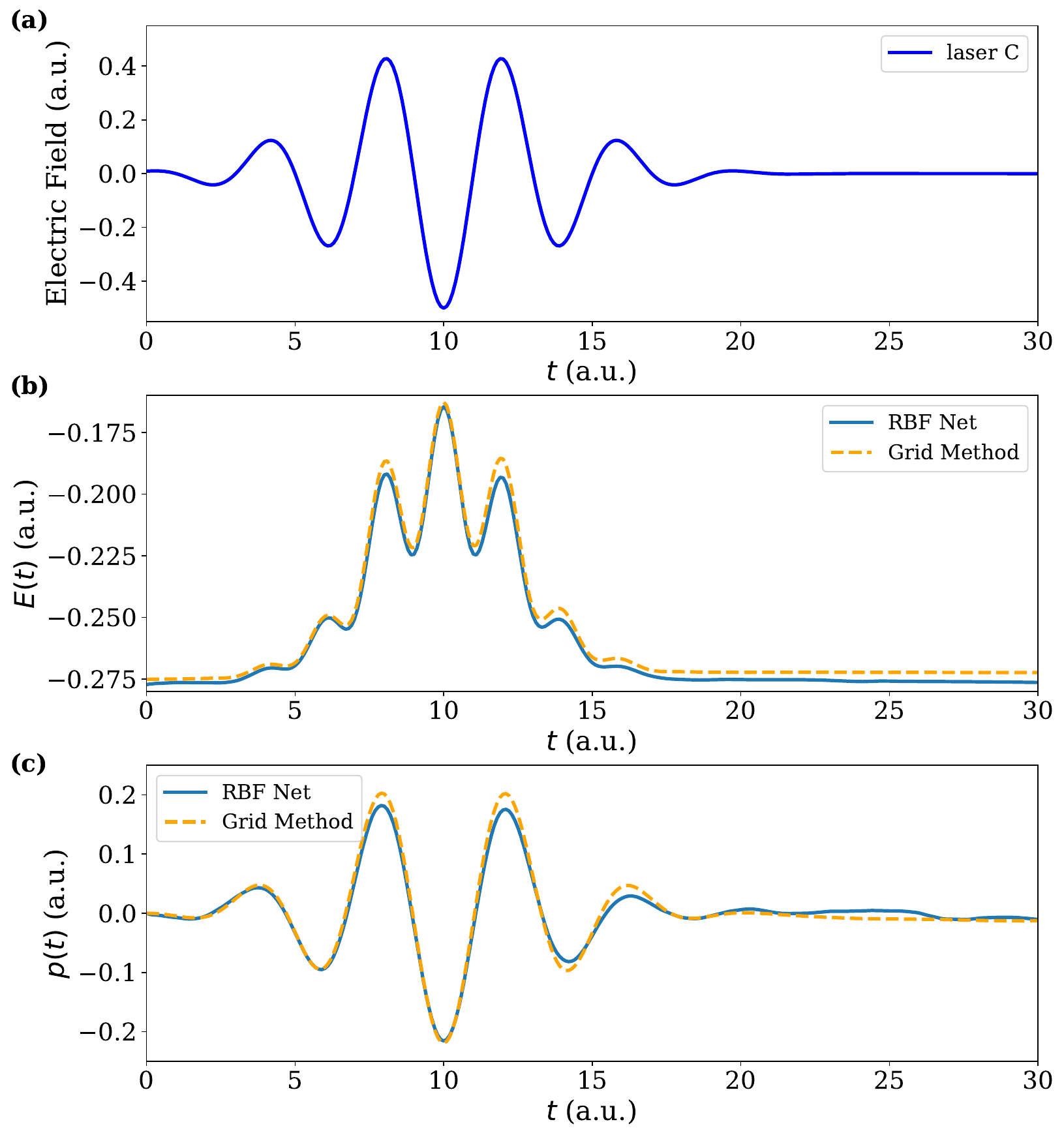}
    \caption{{\bf Simulations of an electron under a soft Coulomb potential in three-dimensional space.} (a) The external electric field. (b) The time evolution of energy. (c) The time evolution of dipole moment.}
    \label{fig:3d}
\end{figure}

To examine the potential of extending our method to realistic systems in the future, here we further study an electron in a three-dimensional soft Coulomb potential subjected to a laser field with parameters $E_0 = 0.5$, $T = 10$, $\tau = 5.0$, and $\omega = \pi/2$.
This field induces moderate excitation without significant ionization, as shown in FIG.~\ref{fig:3d}.
For comparison, we also carry out calculations using the grid method, employing a $40\times40\times40$ spatial mesh with grid spacing $h=0.2$ a.u. and a time step $\Delta t=0.001$ a.u.
Our method achieves very good agreement with the grid method, as shown in FIG.~\ref{fig:3d}(b-c), where the time evolution of energy and dipole moment from both methods align closely.

Due to the large sampling space in 3D and limited computational resources, we disable the adaptive sampling strategy and fix the sample size to $10^4$ throughout the simulation.

\begin{figure}
    \centering       
    \includegraphics[width=0.45\textwidth]{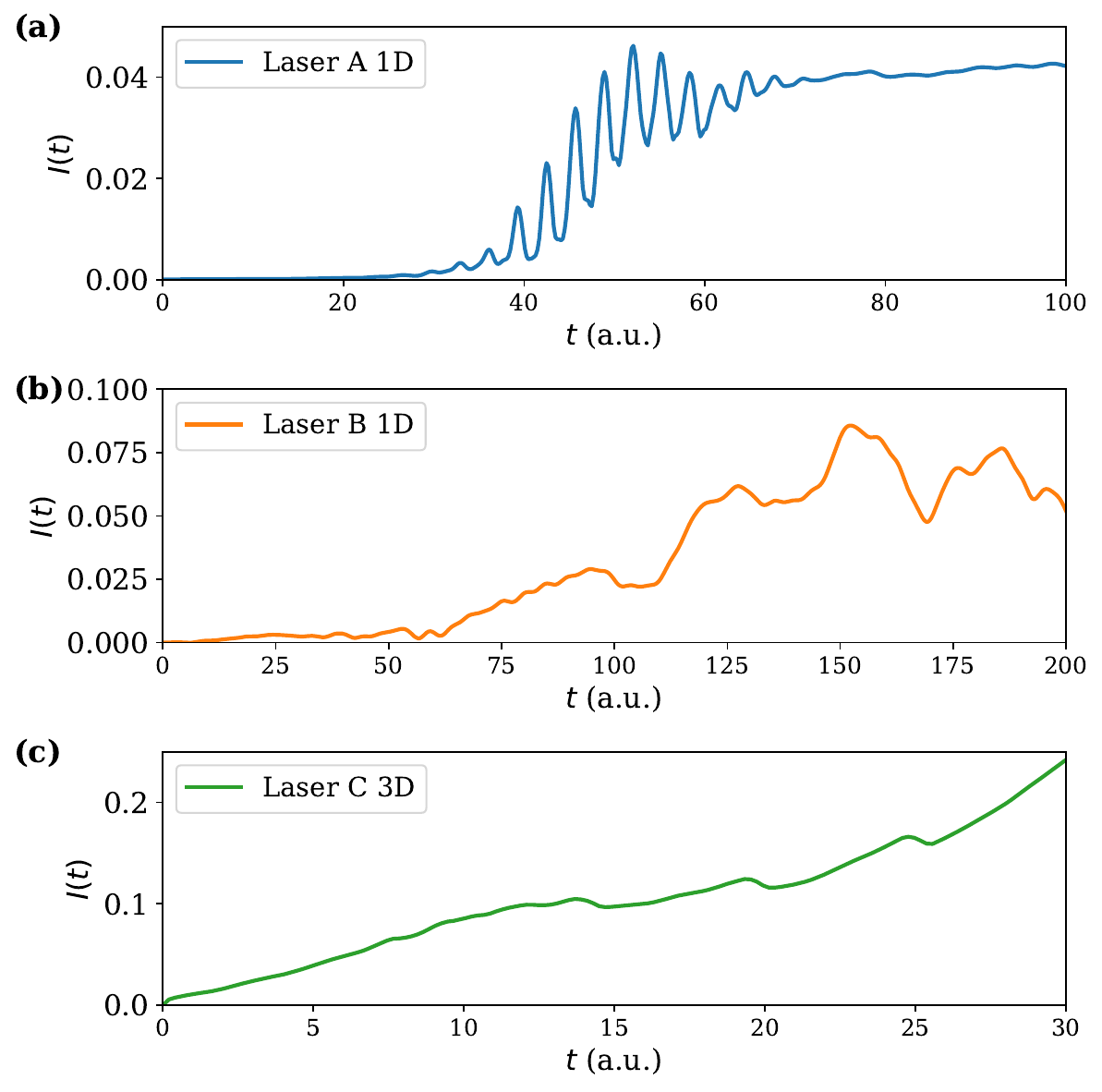}  
    \caption{{\bf Infidelity during simulations under different laser fields with soft Coulomb potential.}}
    \label{fig:infedelity}
\end{figure}

To exploit the cylindrical symmetry of the system, which arises from the laser polarization along the $z$-axis, the network's input features are constructed using cylindrical coordinates $(\rho, z)$, where $\rho=\sqrt{x^2+y^2}$.
This dimensionality reduction improves computational efficiency and enforces symmetry in the learned wavefunction, thereby improving both fitting accuracy and training stability.
Additionally, inspired by the sparse parameter update strategy~\cite{10.5555/3666122.3666301}, we adopt a partial training scheme: during most time steps, only the weights of the output layer are updated, while the parameters of earlier layers are frozen.
This reduces training noise and accelerates convergence.
To prevent overfitting to transient features and maintain long-term stability, all network parameters are unfrozen every 1000 time steps and jointly re-optimized for one time step.
This hybrid update strategy enhances the accuracy and smoothness of wavefunctions over extended simulation durations, thereby improving the stability of the algorithm. However, maintaining long-term accuracy for the 3D case remains a challenge for our approach.
As shown in Fig.~\ref{fig:infedelity}, we use the results of the grid method as the benchmarks and calculate the infidelity~\cite{sinibaldi2023unbiasing} between our method and these benchmarks. The infidelity, $I(t)$, is defined as:
\begin{equation}
    I(t) =1- \frac{|\langle \psi_{\mathrm{RBF}}(t) | \psi_{\mathrm{grid}}(t)\rangle|^2}{{\langle \psi_{\mathrm{RBF}}(t) | \psi_{\mathrm{RBF}}(t)\rangle \cdot \langle \psi_{\mathrm{grid}}(t) | \psi_{\mathrm{grid}}(t)\rangle}},
\end{equation}
where $\psi_{\mathrm{RBF}}$ denotes the RBF network-based wavefunction and $\psi_{\mathrm{grid}}$ denotes the grid method-based wavefunction.
For the 1D cases, the infidelity remains below 0.1 across all tested laser configurations.
Conversely, for the 3D case, the infidelity increases rapidly, indicating a pronounced discrepancy from the benchmark.
Addressing this challenge requires further investigation into higher-dimensional representations and training.


Compared to TDVP-based methods for wavefunction propagation~\cite{PhysRevE.101.023313,Nys2024}, the stochastic representation framework avoids the computational burden of calculating integral matrices and inverting ill-conditioned matrices. Additionally, traditional basis functions or neural networks are typically limited to describing bound states, whereas the RBF layer in our approach enables the representation of scattering states. By dynamically adjusting the coordinates of RBF centers, our method effectively captures wavefunction spreading processes. This makes the RBF network well-suited for ionization simulations.

\section{Conclusion}

We have extended the stochastic representation framework to the real-time domain, leveraging a neural network wavefunction to solve the TDSE. 
By integrating stochastic representation with an adaptive sampling strategy, our method offers two key advantages over conventional grid-based approaches: first, it reduces the number of samples required for accurate wavefunction description; second, it eliminates the need for explicit boundary conditions, which are typically used to absorb emitted wave packets, and thus avoids the necessity of large grids.
In 1D systems, the approach demonstrates high accuracy, particularly in capturing the intricate dynamics of ionization processes under intense laser fields. 
Extending the framework to higher-dimensional systems also shows promising results.
We observe the increased complexity of the wavefunction introduces instability and larger errors, therefore a critical open question is how to ensure the neural network produces a sufficiently smooth wavefunction, as smoothness is essential for obtaining accurate Laplacian values.
Addressing these challenges for more complex systems will require future work, such as integrating advanced machine learning techniques. 
Additionally, incorporating the time-dependent variational principle and natural gradient methods could accelerate parameter optimization, thereby enhancing computational efficiency and facilitating broader applicability to complex quantum systems.

\begin{acknowledgments}
We thank Yubin Qian for valuable discussion.
This work was supported by Beijing Natural Science Foundation under Grant No. JQ22001, the National Key R\&D Program of China under
Grant No. 2021YFA1400500, and National Science Foundation of China under Grant No. 12334003. 
We are grateful for computational resources provided by the High Performance Computing Platform of Peking University.
\end{acknowledgments}

\begin{appendices}
\section{Integrator Stability Analysis}
\label{sec:inte}
\setcounter{equation}{0}
\renewcommand\theequation{A\arabic{equation}}
\begin{figure}[ht]
  \centering
  \includegraphics[width=0.45\textwidth]{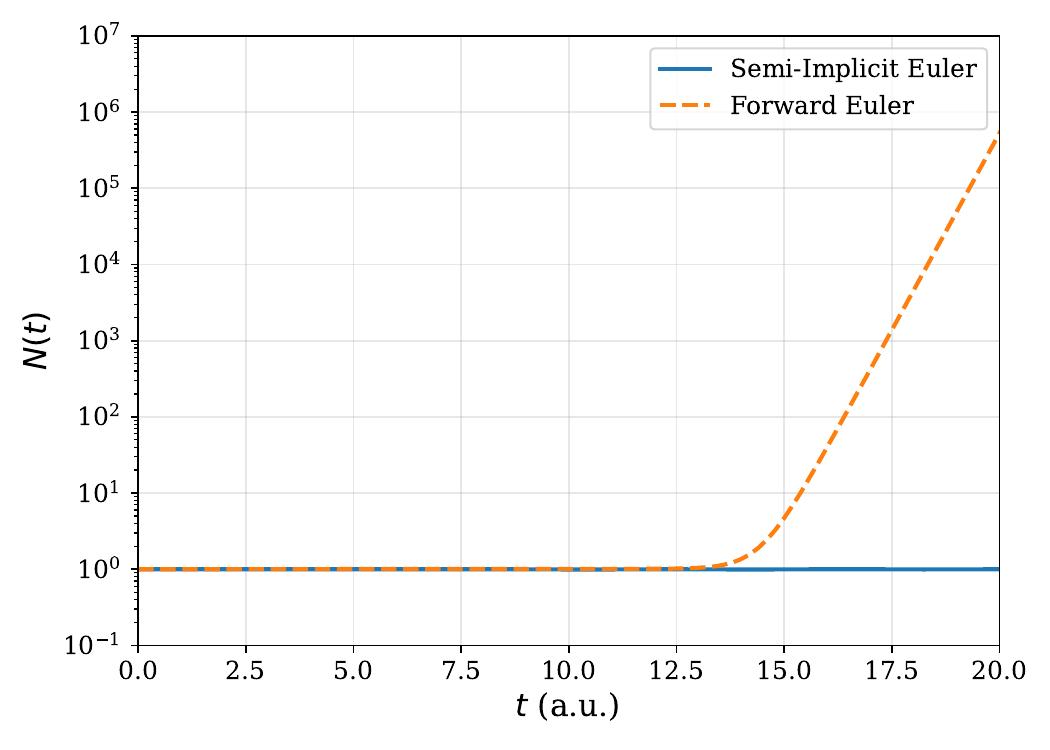}
  \caption{{\bf Comparison of the wavefunction norm  evolution.}  Norm evolution under Laser A, soft Coulomb potential. The forward Euler method exhibits exponential growth in norm, while the semi-implicit Euler method maintains norm conservation.  The semi-implicit method's symplectic nature ensures long-term stability.}
  \label{fig:euler_comparison}
\end{figure}
Here we provide an analysis of the numerical integrator used in the main text.
During the time evolution, the wavefunction evolves as
\begin{equation}
    \psi^{(n+1)} = e^{-i\Delta t\hat{H}}\psi^{(n)}\approx(1-i\Delta t\hat{H})\psi^{(n)}.\label{eq:evolve}
\end{equation}
As in the main text, we decompose the wavefunction into its real and imaginary parts, $\psi^{(n)} = u^{(n)} + iv^{(n)}$. Eq.~\ref{eq:evolve} then yields a forward Euler method:
\begin{equation}
    \left\{\begin{array}{c}
        u^{(n+1)} = u^{(n)} + \Delta t \hat{H}v^{(n)},  \\
        v^{(n+1)} = v^{(n)} - \Delta t \hat{H}u^{(n)}.
    \end{array}\right.
\end{equation}
However, in this work we adopt the semi-implicit Euler method instead, which reads
\begin{equation}
    \left\{\begin{array}{c}
        u^{(n+1)} = u^{(n)} + \Delta t \hat{H}v^{(n)},  \\
        v^{(n+1)} = v^{(n)} - \Delta t \hat{H}u^{(n+1)}.
    \end{array}\right.
\end{equation}
Both the forward Euler and semi-implicit Euler methods have a local truncation error of $\mathcal{O}(\Delta t^2)$, meaning they are both first-order accurate globally. The critical difference lies in their long-term stability, which is determined by their geometric properties. We can analyze this by writing the update as a matrix transformation on the vector $\mathbf{z}^{(n)} = (u^{(n)}, v^{(n)})^T$ and let $\hat{K} = \Delta t \hat{H}$.

For the forward Euler method, the transformation is:
\begin{equation}
    \begin{pmatrix} u^{(n+1)} \\ v^{(n+1)} \end{pmatrix} = 
    \begin{pmatrix} 1 & \hat{K} \\ -\hat{K} & 1 \end{pmatrix}
    \begin{pmatrix} u^{(n)} \\ v^{(n)} \end{pmatrix} = M_{\mathrm{E}} \mathbf{z}^{(n)}.
\end{equation}
The determinant of this transformation matrix is $\det(M_{\mathrm{E}}) = 1 + \hat{K}^2 > 1$. A transformation with a determinant greater than 1 expands the phase-space volume at each step. For this system, this manifests as a systematic drift in the wavefunction norm ($||\psi||^2 = ||u||^2 + ||v||^2$), causing the solution to become unstable and grow without bound.

For the semi-implicit Euler method, the transformation is:
\begin{align}
    u^{(n+1)} &= u^{(n)} + \hat{K} v^{(n)}, \\
    v^{(n+1)} &= v^{(n)} - \hat{K} u^{(n+1)} = -\hat{K} u^{(n)} + (1 - \hat{K}^2) v^{(n)}.
\end{align}
Writing this in matrix form:
\begin{equation}
    \begin{pmatrix} u^{(n+1)} \\ v^{(n+1)} \end{pmatrix} = 
    \begin{pmatrix} 1 & \hat{K} \\ -\hat{K} & 1 - \hat{K}^2 \end{pmatrix}
    \begin{pmatrix} u^{(n)} \\ v^{(n)} \end{pmatrix} = M_{\mathrm{SI}} \mathbf{z}^{(n)}.
\end{equation}
The determinant of this matrix $M_{\mathrm{SI}}$ is $\det(M_{\mathrm{SI}}) = 1 - \hat{K}^2 + \hat{K}^2 = 1$.
A transformation with a determinant of 1 is symplectic (or more precisely for this linear system, volume-preserving). This property is the defining feature of Hamiltonian dynamics, which the Schrödinger equation follows. The semi-implicit integrator, by preserving this geometric property, also preserves the norm of the wavefunction. It does not suffer from the systematic drift like the forward Euler method, allowing for stable, long-term integrations that remain bounded, which we find to be the case in practice.

This theoretical difference in stability is demonstrated in Fig.~\ref{fig:euler_comparison}, which shows a comparison of the wavefunction norm over time when simulated with both the forward Euler and the semi-implicit Euler methods. The forward Euler method clearly shows an exponential drift, leading to a catastrophic failure in norm preservation. In contrast, the semi-implicit method's norm remains correct, consistent with its symplectic nature.

While the semi-implicit method provides excellent stability at a low computational cost, it is still a first-order method. For applications requiring higher accuracy (i.e., a smaller local truncation error), one could employ a second-order symplectic integrator, such as the velocity Verlet algorithm~\cite{martys1999velocity}.

\section{Sensitivity Analysis of MAE threshold}
\label{sec:mae}

\begin{figure}[ht]
  \centering
  \includegraphics[width=0.45\textwidth]{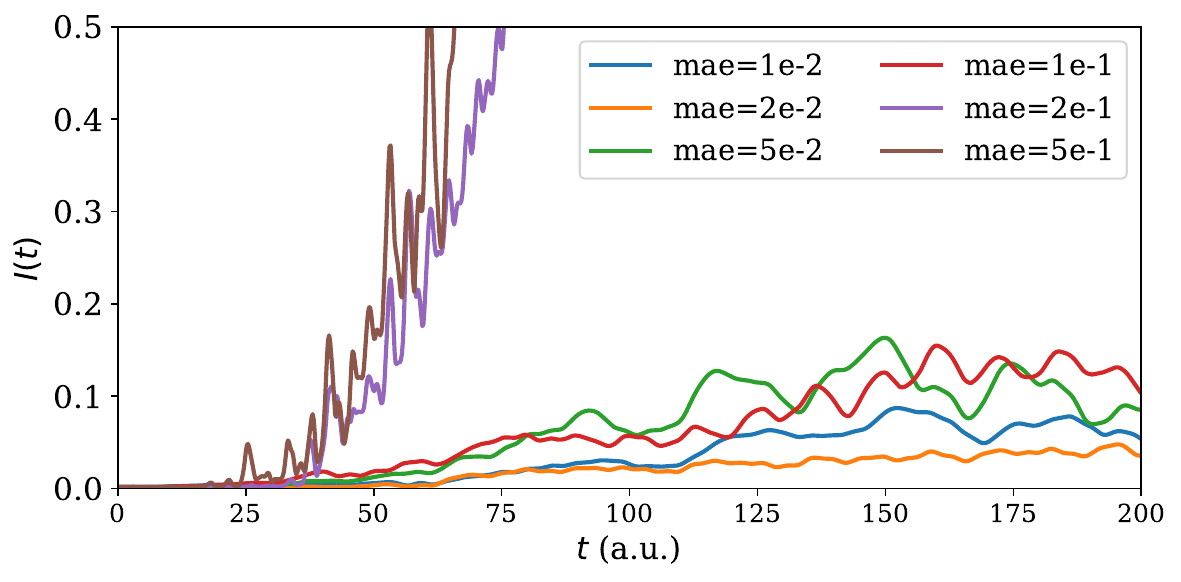}
  \caption{{\bf Infidelity during simulations under different MAE threshold \(\delta\), 1D soft Coulomb potential, Laser B}}
    \label{fig:infedelity_mae}
\end{figure}

Here we examine how the MAE threshold \(\delta\) influences the fidelity decay during simulation. Fig.~\ref{fig:infedelity_mae} displays the infidelity trajectories under 6 different MAE thresholds. For \(\delta<0.1\), our method remains robust and stable, with \(\delta=0.02\) achieving the best performance. For \(\delta>0.1\), the overly large threshold leads to underfitting, causing the infidelity to rise rapidly.

\end{appendices}

\bibliographystyle{unsrt}
\bibliography{references}

\end{document}